\documentclass[twocolumn]{aastex631}

\shorttitle{Peculiar Dust Emission in OMC 2/3}
\shortauthors{Nozari et al.}
\graphicspath{{./}{figures/}}
\newcommand{\Jybeam}{\text{Jy beam$^{-1}$}} 
\usepackage{graphicx}
\usepackage{natbib}
\usepackage{txfonts}
\usepackage[super]{nth}
\begin{document}

\title{Peculiar Dust Emission within the Orion Molecular Cloud}

\author{Parisa Nozari}
\affiliation{Queen's University, 64 Bader Ln, Kingston, ON K7L3N6, Canada}

\author{Sarah Sadavoy}
\affiliation{Queen's University, 64 Bader Ln, Kingston, ON K7L3N6, Canada}

\author{Edwige Chapillon}
\affiliation{The Institut de radioastronomie millimétrique (IRAM), Grenoble, France}

\author{Brian Mason}
\affiliation{National Radio Astronomy Observatory, 520 Edgemont Road, Charlottesville, VA 22903, USA}

\author{Rachel Friesen}
\affiliation{Department of Astronomy \& Astrophysics, University of Toronto, 50 St. George St., Toronto, ON M5S 3H4, Canada}

\author{Ian Lowe}
\affiliation{ Department of Astronomy and Steward Observatory, University of Arizona}

\author{Thomas Stanke}
\affiliation{Max Planck Institute for Extraterrestrial Physics, Garching bei M\"{u}nchen, Germany}

\author{James Di Francesco}
\affiliation{NRC Herzberg Astronomy and Astrophysics, 5071 West Saanich Road, Victoria, BC V9E 2E7, Canada}

\author{Thomas Henning}
\affiliation{Max Planck Institute for Astronomy, K\"{o}nigstuhl 17, D-69117 Heidelberg, Germany}

\author{Qizhou Zhang}
\affiliation{Center for Astrophysics, Harvard \& Smithsonian, 60 Garden Street, MS 42, Cambridge,
MA 02138, USA}

\author{Amelia Stutz}
\affiliation{Departamento de Astronom\'{i}a, Universidad de Concepci\'{o}n, Casilla 160-C, Concepci\'{o}n, Chile}

\begin{abstract}
It is widely assumed that dust opacities in molecular clouds follow a  power-law profile with an index, $\beta$. Recent studies of the Orion Molecular Cloud (OMC) 2/3 complex, however, show a flattening in the spectral energy distribution (SED) at $ \lambda > 2$ mm implying non-constant indices on scales $\gtrsim$ 0.08 pc. The origin of this flattening is not yet known but it may be due to the intrinsic properties of the dust grains or contamination from other sources of emission. We investigate the SED slopes in OMC 2/3 further using observations of six protostellar cores with NOEMA from 2.9 mm to 3.6 mm and ALMA-ACA in Band 4 (1.9 -- 2.1 mm) and Band 5 (1.6 -- 1.8 mm)    on core and envelope scales of $\sim 0.02 - 0.08$ pc. We confirm flattened opacity indices between 2.9 mm and 3.6 mm for the six cores with    $\beta \approx -0.16 - 1.45$, which are notably lower than the $\beta$ values of $> 1.3$ measured for these sources on $0.08$ pc scales from single-dish data. Four sources have consistent SED slopes between the ALMA data and the NOEMA data.  We propose that these sources may have a significant fraction of emission coming from large dust grains in embedded disks, which biases the emission more at longer wavelengths. Two sources, however, had inconsistent slopes between the ALMA and NOEMA data, indicating  different origins of emission.
These results highlight how care is needed when combining multi-scale observations or extrapolating single-band observations to other wavelengths.
\end{abstract}


\section{Introduction}

The OMC 2/3 region within the Orion Molecular Cloud (OMC) complex contains a rich, nearby filamentary structure that enables detailed observations of various stages of star formation processes, making it an invaluable laboratory for studying the fundamental physics governing the birth of stars and planetary systems \citep[e.g.,][]{Chini1997,Peterson2005,Sadavoy2010,Stutz2015,Salji2015,Megeath2016,Hacar2018}. The proximity and wealth of the young stellar content in OMC 2/3 allows researchers to map its star formation activity and connect that activity to the properties of the filament.  

Thermal dust emission is particularly useful as a tracer of molecular cloud mass and structure, despite gas being more abundant in the interstellar medium (ISM). First, thermal dust emission can be used to derive temperature and (column) density information. Second, dust is found on all scales, whereas emission from a single molecular species will trace gas at, e.g., a narrow range of densities, and thus will be sensitive to a limited range of physical scales \citep[e.g.,][]{Caselli2002,Crapsi2005,Ceccarelli2007,Andersson2015,Guillet2020,Liu2024}.

Dust opacity is an intrinsic property of dust grains that specify their ability to absorb light and ultimately re-radiate light. It is commonly described by a power-law profile of the form $\kappa_\nu \propto \nu^\beta$ at long wavelengths ($\lambda > 100 \ \mathrm{\mu m}$), but the slope of this power law varies with dust properties, particularly grain size \citep[e.g.,][]{Ossenkopf1994,D'Alessio2001, Ormel2011, Testi2014}. Nevertheless, many studies characterize dust opacity with fixed values of $\beta$. For instance, values of $\beta \approx 1.8 - 2$ are often used for dust opacity in molecular clouds and dense cores \citep[e.g.,][]{Ysard2012,juvela2018,Long2020}, whereas values of $\beta < 1$ are typical for protoplanetary disks \citep[e.g.,][]{Birnstiel2012,Andrews2018,Hendler2020,Tazzari2021}.

Contrary to these expectations, multiple recent studies on the OMC 2/3 complex have reported conflicting results regarding the $\beta$ index. \citet{Schnee2014} combined 3 mm continuum observations from the Green Bank Telescope (GBT) and 1 mm data from the IRAM 30 m Telescope to measure consistently low values for $\beta$ of $\sim 0.9 \pm 0.3 $ on scales of 0.02 -- 0.1 pc toward the dense cores and filaments of the OMC 2/3.  These low values are more similar to what is measured in disks, whereas they were measuring $\beta$ on core-scales and in filaments where steeper values are expected. Using \emph{Herschel} data at $160-500 \, \mathrm{\mu m}$ and IRAM 30 m data at 2 mm to examine the larger-scale cloud, \citet{Sadavoy2016} found more typical values of $\beta$ in OMC 2/3. Their SED fits gave $\beta > 1.2$ across all of OMC 2/3, with most of the filament exhibiting $\beta \approx 1.7-1.8$ on scales greater than 0.08 pc.  Instead, their SEDs suggested that the SED flattens at $\lambda > 2$ mm.  More recently, \citet{mason2020} confirmed the flatter spectral index at $\lambda > 2$ mm in OMC 2/3 using new 3 mm and 1 cm data from the GBT with \emph{Herschel} observations.  Specifically, they examined SED using dust emission from slices through the filament, demonstrating elevated emission at $\lambda \gtrsim 3$ mm even toward starless cores and filamentary material on scales of $\sim 0.08$ pc. 

While studies such as \citet{Miotello2014}, \citet{Li2017}, and \citet{Galametz2019} have identified $\beta \sim 1$ in the envelopes of Class 0 young stellar objects (YSOs), these measurements are typically on scales $< 0.01$ pc ($<$ 2000 au), whereas the aforementioned OMC 2/3 studies find these values of $\beta$  over a much larger spatial scale of $\sim 0.08 - 1$ pc.  Thus, OMC 2/3 shows a discrepancy in spectral slope on core scales. Moreover, following the approach by \citet{mason2020}, \citet{Lowe2022} showed similar flattened spectral slopes in slides through Orion A, Orion B, and Serpens, indicating that this discrepancy is not unique to OMC 2/3 and must be commonplace. Normally, flatter SEDs at long wavelengths can be indicative of different physical processes, such as free-free emission or Anomalous Microwave Emission (AME) \citep[e.g.,][]{Draine1998,Draine2011,Planck2016_diffuseforeground}.  Nevertheless, neither of these processes appear to explain the flattened SED profiles \citep{mason2020, Lowe2022}. Free-free emission is typically associated with protostellar jets or winds \citep[e.g.][]{Anglada1995, Anglada1998, shang2004, Dicker2009} and is normally not typically seen on core scales $\gtrsim 0.1$ pc or toward starless cores and filaments.  AME is attributed to small dust grains that are spinning and is mostly detected on much larger (cloud) scales \citep[e.g.,][]{Ysard2011, Hoang2016, mason2020} than what has been studied.

To characterize the change in SED slope seen in OMC 2/3 better, we obtained Northern Extended Millimeter Array (NOEMA) data at 3 mm and Atacama Compact Array (ACA) Band 4 (137-153 GHz) and Band 5 (163-179 GHz) observations for six, bright protostellar cores in OMC 2/3 on $< 0.1$ pc scales. These cores all exhibited elevated 3 mm emission based on prior \emph{Herschel} and GBT data \citep{Sadavoy2016, mason2020}, and they are all bright, compact, and isolated, making them the best targets to map the SED slope by imaging the basebands separately and to avoid confusion with extended emission or neighboring sources. By comparing multiwavelength data on $<0.1$ pc scales, we aim to identify how the SED slope is changing over the critical wavelength range between 1 and 3 mm using data that capture similar scales.
 
This paper is structured as follows: Section 2 presents the NOEMA and ALMA observations; Section 3 discusses the analysis and slope calculations; Section 4 presents the results; Section 5 discusses the individual sources; Section 6 explores several physical mechanisms that could explain the SED properties; finally, Section 7 summarizes the key results and interpretations, and presents our conclusions.

\section{Observations} \label{sec:obs}

\subsection{NOEMA Data}

We observed six dense cores within the OMC 2/3 filament with NOEMA in D-configuration with 10 antennas on 19 February 2019. Table \ref{tab:list of sources} lists the six cores and their properties. The correlator was set to 81-104 GHz, with a total bandwidth of 16 GHz split between four basebands with 2 MHz resolution and 1828 channels per baseband. Additionally, 11 smaller windows were included for higher spectral resolution, but these are not examined here. The total observing time was 1.18 hours with single pointings over a shared track. The observations used 3C84 for flux calibration and B0458-020 and B0539-057 for phase calibration. Phase and amplitude radio frequency (RF) calibration was performed by interpolating between channels, and a spectral index of -0.38 was adopted for the RF calibrator.

To reduce effects from unstable weather conditions, initial scans were flagged to ensure accurate calibration unaffected by transient phase and amplitude fluctuations. Calibration was performed primarily on Antenna 1 before extending the solution to all receivers, ensuring consistency. Our NOEMA calibration involved both frequency and time domains, correcting instrumental effects through standard 30-80 degree phase thresholds for acceptable system deviation. The standard pipeline executed bandpass calibration and indicated consistent antenna response. The calibration uncertainties were estimated to be $<$ 40 degrees in phase rms,  $<$ 21.6\% amplitude loss, $<$ 30\% in pointing error, and $<$ in 30\% focus error. After calibration, molecular line subtraction was performed before imaging. Lines were subtracted per source and baseband by flagging line channels. A filtering process then extracted line emission, generating a continuum UV-table for dirty image creation prior to cleaning.

\begin{deluxetable*}{cccccccc}
%\tablenum{1}
\tablecaption{OMC 2/3 NOEMA Targets and Information}\label{tab:list of sources}
\tablewidth{0pt}
\tablehead{
\colhead{Source} & \colhead{$\alpha$} & \colhead{$\delta$ } &
\colhead{Class} & \colhead{M$_{disk, ALMA}$} & \colhead{M$_{disk, VLA}$} & \colhead{M$_{env}$} & \colhead{HOPS} \\[-1.5mm]
\colhead{} & \colhead{(J2000)} & \colhead{(J2000)} &
\colhead{} & \colhead{($M_{\oplus}$)} & \colhead{($M_{\oplus}$)} & \colhead{($M_{\odot}$)} &\colhead{}\\[-8mm]
}
\decimalcolnumbers
\startdata
FIR2 & 05:35:24.305 & -05.08.30.58 & I & 58.4 & 403.1 & 0.96 &  {HOPS-68}\\
FIR6B & 05:35:23.289 & -05.12.03.72 & 0 & 152.7 & 790.3 &0.11 &  {HOPS-60}\\ 
MMS6 & 05:35:23.442 & -05.01.30.44 & 0 & 627.5 & 1510.3 & 0.097&  {HOPS-87}\\
MMS7 & 05:35:26.545 & -05.03.55.28 & I & 168.4 & 700.8 & 0.034&  {HOPS-84}\\
MMS9 & 05:35:25.993 & -05.05.43.90 & 0 & 10.4 & 106.4 & 0.24 &  {HOPS-78} \\
NW167 & 05:35:29.720 & -04.58.48.86 & 0 & 135.0 & 759.0 &1.9 &  {HOPS-96} \\
\enddata
\tablecomments{Source names (column 1) are based on \citet{Chini1997}. Protostellar classifications (column 4) and disk masses (columns 5 and 6) are taken from \citet{Tobin2020}, which presents ALMA 0.87 mm and VLA 9 mm observations. The disk masses represent the summed total of all disks located within the NOEMA beam ($< 6\arcsec$) at the position of each source. The envelope masses (column 7) are taken from the \emph{Herschel} Orion Protostellar Survey \citep[HOPS,][]{Furlan2016}.  {The corresponding HOPS name of each source is given in column 8.}}
\end{deluxetable*}

% table about the observations

\begin{deluxetable*}{cccccccccc}
\label{obs_table}
%\tablenum{2}
\tablecaption{Resolution and Sensitivity for NOEMA and ALMA}\label{NOEMA_table}
\tablewidth{0pt}
\tablehead{
\colhead{Baseband Id} & 
\colhead{$\nu$\tablenotemark{a}} & 
\colhead{$\lambda$} & 
\colhead{$a$} &
\colhead{$b$} & 
\colhead{PA} & 
\colhead{rms\tablenotemark{b}} &
\colhead{$a$\tablenotemark{c}} &
\colhead{$b$\tablenotemark{c}} & 
\colhead{PA\tablenotemark{c}}\\[-1.5mm]
\colhead{} & 
\colhead{(GHz)} & 
\colhead{(mm)} & 
\colhead{(\arcsec)} &
\colhead{(\arcsec)} & 
\colhead{(deg)} & 
\colhead{($\mathrm{m}\Jybeam$)} & 
\colhead{(\arcsec)} &
\colhead{(\arcsec)} & 
\colhead{(deg)}\\[-8mm]
}
\decimalcolnumbers
\startdata
\multicolumn{10}{c}{NOEMA Data}\\
\hline
LSB-O & 82.71897 &  {3.63} & 7.0 & 4.3 & 26.2 & 0.11 &  {14.2} &  {8.8} &  {87.1}\\
LSB-I & 86.78157 &  {3.46} & 6.3 & 4.1 & 26.1 & 0.23 &  {13.2} &  {8.4} &  {86.3}\\
USB-I & 98.20925 &  {3.05} & 6.0 & 3.6 & 25.6 & 0.27 &  {11.4} &  {7.9} &  {-85.1}\\
USB-O & 102.2718 &  {2.93} & 5.7 & 3.5 & 25.0 & 0.31 &  {10.7} &  {7.6} &  {-86.0}\\ 
\hline
\multicolumn{10}{c}{ALMA Data\tablenotemark{d}}\\
\hline
Band 4 - Low & 138.9607 &  {2.16} & 13.7 & 5.7 & -87.7 & 1.1 &  {14.2} &  {8.9} &  {86.7}\\
Band 4 - High & 150.9918 &  {1.99} & 12.2 & 5.7 & -87.9 & 1.5 &  {13.2} &  {8.4} &  {86.2}\\
Band 5 - Low & 164.9749 &  {1.82} & 10.8 & 5.4 & -80.8 & 1.2 &  {11.5} &  {7.9} &  {-79.4}\\
Band 5 - High & 177.007 &  {1.69} & 9.7 & 5.0 & -81.4 & 1.7 &  {10.7} &  {7.6} &  {-79.1}\\
\enddata
\tablenotetext{a}{Central frequency of each baseband.  NOEMA basebands are 4 GHz, ALMA basebands are 3.6 GHz.  The ALMA data combine the two lower sideband spectral windows (Low) and two upper sideband spectral windows (High) for better sensitivity. }
\tablenotetext{b}{The rms levels in the table are relatively consistent between all sources at each frequency band.  For the ALMA data, the rms values will be 1.2 times higher for MMS9 and NW167 because of their primary beam corrections. }
\tablenotetext{c}{Final resolution after smoothing the NOEMA and ALMA data using the beam of the other data. We paired the ALMA and NOEMA data as follows: LSB-O with Band 4-Low, LSB-I with Band 4-High, USB-I with Band 5-Low, and USB-O with Band 5-High.}
\tablenotetext{d}{All values are for a uv $> 5~\mathrm{k\lambda}$ cut.}
\end{deluxetable*}

\begin{figure*}[h!]
\includegraphics[width=\textwidth,height=9cm]{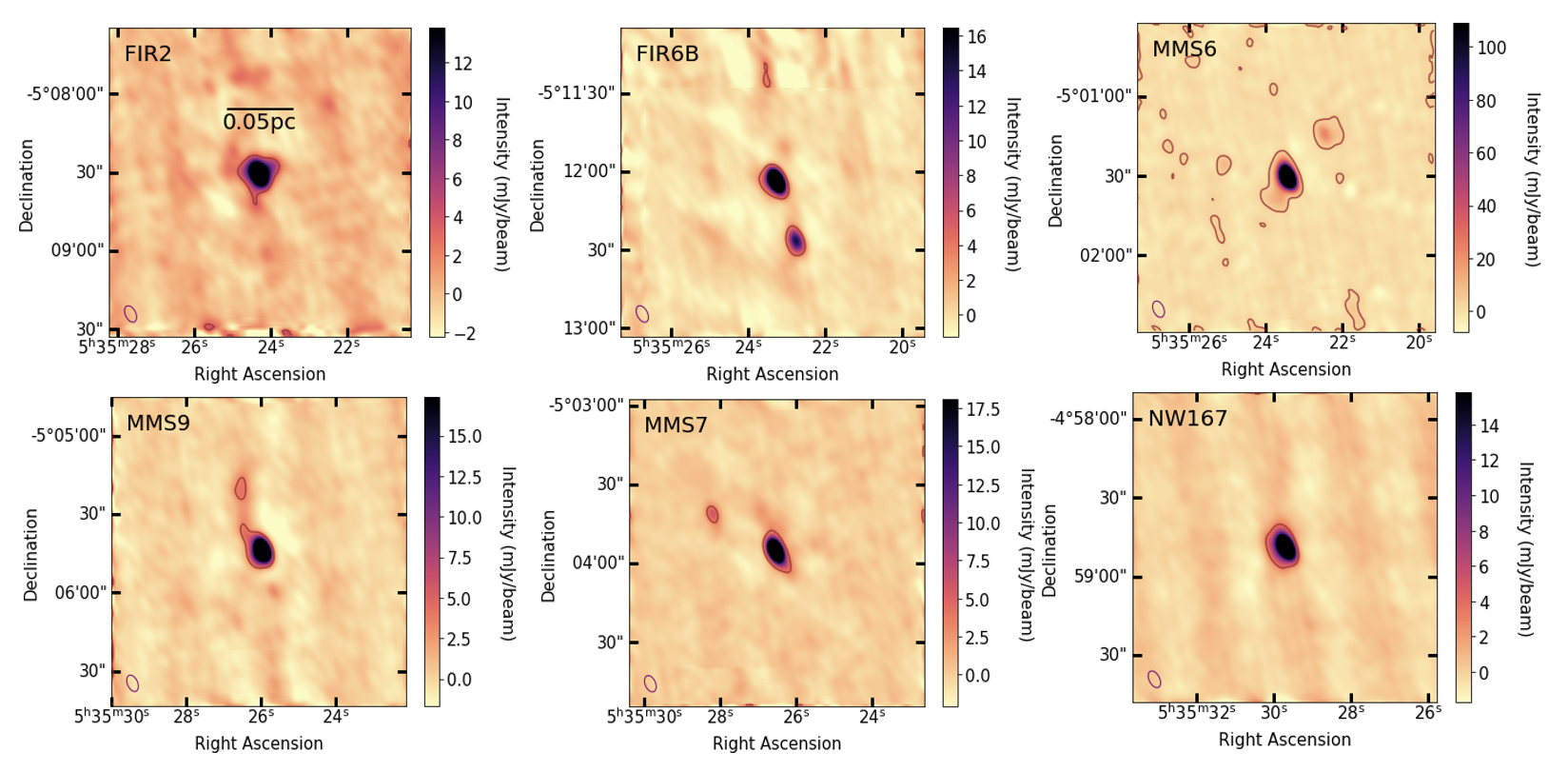}
\caption{Cleaned NOEMA images of all six protostellar cores in OMC 2/3 for the LSB-I baseband at a central frequency of 86.78157 GHz.   {Contours represent a 5$\sigma$ level.} The synthesized beam is plotted in the bottom left corner.   {These maps represent the unsmoothed NOEMA observations.}}
\label{fig:source_gallery}
\end{figure*}

The NOEMA visibilities were cleaned using the Hogbom deconvolution algorithm \citep[e.g.,][]{Hogbom1974} with 200 iterations and a noise level of 2.5 m\Jybeam\ on the dirty images. To further enhance image accuracy, phase only self-calibration was conducted as the target signal-to-noise ratios (SNRs) were sufficiently high ($>100$).  We performed one round of phase only self-calibration with interval of 45s. Subsequently, another round of cleaning was carried out with a reduced noise threshold of 0.2 m\Jybeam. We employed the \texttt{MAPPING} method in GILDAS \citep[e.g.,][]{GILDAS} for data imaging with a map size of 128 by 128 pixels, corresponding to 92.16\arcsec\ by 92.16\arcsec, and 0.72\arcsec\ pixels. Since the observations had such high SNR, we imaged each of the four basebands separately to infer the SED slopes better.  {Table \ref{obs_table} gives the frequency, resolution, and sensitivity for each NOEMA baseband at the native resolution (columns 3 -- 5) and for the smoothed data used in our analysis (columns 7 -- 9).} Figure \ref{fig:source_gallery} shows images of all our targets at 86.78 GHz, corresponding to the inner lower side band (LSB-I). The targets are singular except for FIR6B where a second compact source is seen south of the main target. Several sources also show extended emission like MMS9 and FIR2. The measured properties of each compact source are detailed in Section 3.

\subsection{ALMA-ACA Data}

We observed the entire OMC 2/3 filament in Band 4 (137-153 GHz) and Band 5 (163-179 GHz) with the Atacama Large Millimeter/submillimeter Array (ALMA) ACA from October 2017 to December 2017. The region was divided into two mosaics with one centered on OMC 2 (area of 3.2 $\times$ 7.0 arcmin) and the other on OMC 3 (area 3.7 $\times$ 7.5 arcmin in Band 4 and 3.4 $\times$ 7.4 arcmin in Band 5) for a total observing time of 9.53 hours. The Band 4 and Band 5 datasets were calibrated through standard procedures from the North American ALMA Science Center (NAASC). The flux calibrator was J0522-3627, and the phase calibrator was J0541-0211 for the Band 4 data and J0542-0913 for the Band 5 data.

\begin{figure*}[h!]
\includegraphics[trim=1.0cm 2mm 1mm 1mm,clip=true,width=0.97\textwidth]{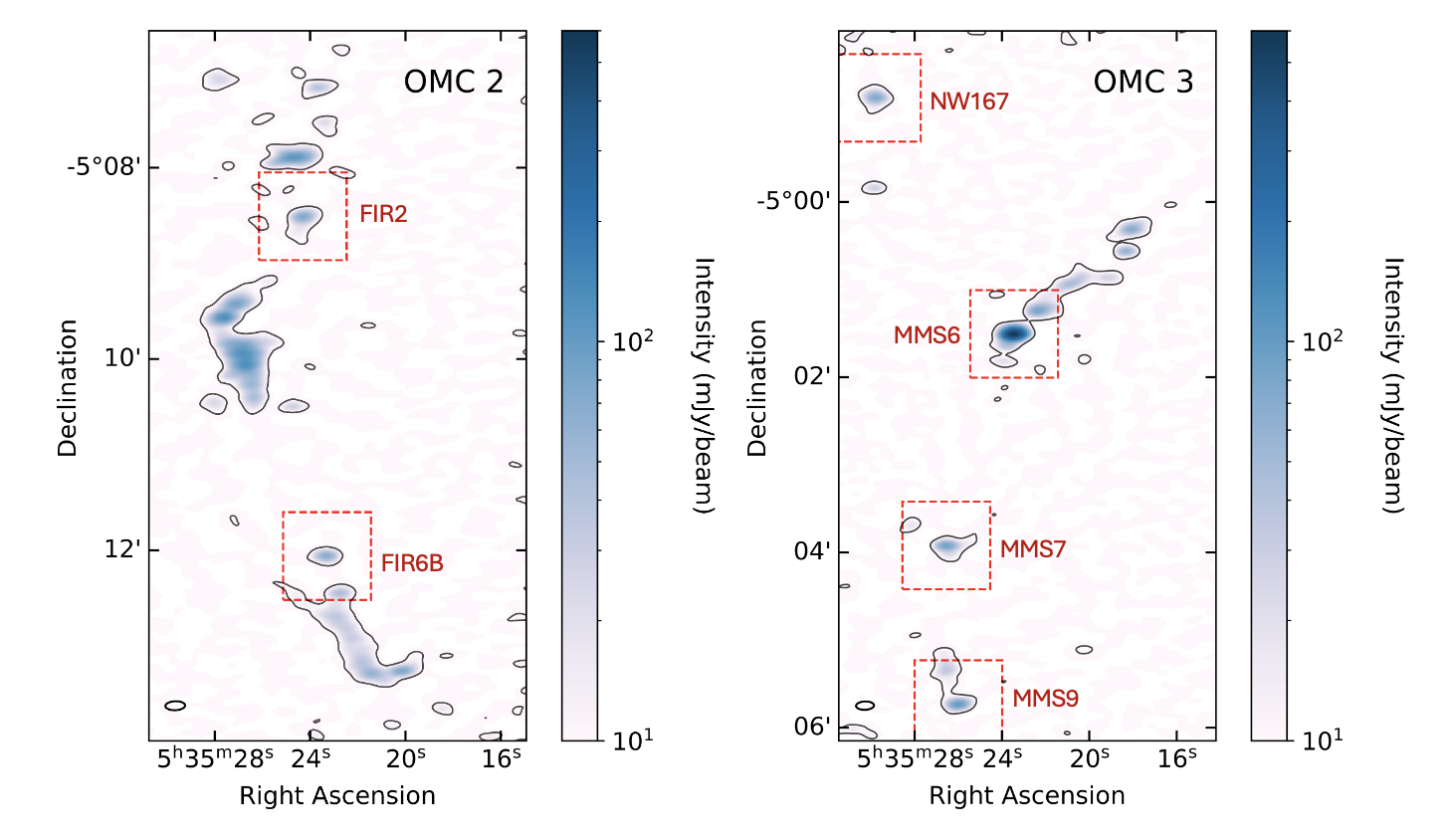}
\caption{Cleaned continuum emission maps of OMC 2 (left) and OMC 3 (right) at Band 4-High (150.9918 GHz) from ALMA/ACA observations as an example.  {These maps correspond to the unsmoothed mosaics with a uv-cut of 5 k$\lambda$.  A 5$\sigma$ contour is overlaid on the maps.} The six target sources are labeled and the synthesized beam is shown in the lower-left corner.  {The red boxes represent the regions used for flux extraction with \texttt{imfit} in CASA.}}
\label{figomcmap}
\end{figure*}

We employed the \texttt{tclean} algorithm in the CASA software \citep{McMullin2007}, utilizing the Hogbom deconvolver. Prior to cleaning, the ALMA data were inspected to identify any channels that could be contaminated by line emission and such channels were flagged.
We applied an iterative approach for phase-only self-calibration, similar to the method described by \citet{Ginsburg2018}.  {We applied three rounds of phase only self-calibration with successively smaller solution intervals starting from the full scan to 30s and then 15s.} For all cleaning rounds, we used Briggs weighting with a robust parameter of 0.5 to optimize both sensitivity and resolution \citep{Briggs1995,Briggs1999}. Since our targets are bright, we imaged each sideband separately to obtain a better measure of the spectral index between 136 GHz and 177 GHz.   {As a result, we have two measurements for each of Band 4 and Band 5 using the baseband pairs, where the paired lower basebands are ``Low'' and the paired upper basebands are ``High''.}

To ensure a consistent comparison between the ALMA and NOEMA data, we generated images from the ACA visibilities with a uv range cutoff of uv $> 5~k\lambda$, which corresponds to spatial scales of $\lesssim$ 40 arcsec. This choice was motivated by the minimum baseline length of approximately 5 k$\lambda$ observed in our NOEMA data. The purpose of specifying the uv range was to enable the characterization of the SED slope for each target using data that trace the same spatial scales. Table \ref{obs_table} gives the ALMA synthesized beam information and map sensitivities for the ``Low'' and ``High'' Band 4 and 5 data following this uv cut (columns 4 -- 6) and the smoothed data used in the following analysis (columns 7 -- 9). Figure \ref{figomcmap} shows the clean Band 4 mosaic maps of OMC 2 and OMC 3 as an example.

\section{Methods} \label{sec:methods}

\subsection{Source Extraction} \label{subsec:Flux-est}

To construct SEDs for each source, we need the total fluxes across the multiple datasets.  While we used a uv-cut to ensure that the ALMA-ACA data and NOEMA data cover similar spatial scales, the resulting synthesized beams differ by factors of 2.5 -- 2.7 in area.  We therefore smoothed the datasets to better match in resolution.  For simplicity, we smoothed the ALMA and NOEMA data by the synthesized beam of the other, matching the highest to lowest frequencies for each pair (e.g., LSB-O with Band 4-Low, LSB-I with Band 4-High, etc.).  The resulting smoothed beam resolutions are given in columns 7 -- 9 in Table \ref{obs_table}.  Therefore, the smoothed maps consistently cover scales from 0.02 pc -- 0.08 pc.  We stress that unlike prior studies which combined data with different resolutions and different degrees of filtering, these observations should be consistently tracing emission from the envelope and core over the same spatial scales.

The integrated flux of each source is extracted using the \texttt{imfit} task in CASA. This task fits the source emission with two-dimensional Gaussian functions in the image plane, allowing us to estimate source parameters  {such as the position, size, and total flux of each source.} For the NOEMA data, we used a $30\times30$ pixel box centered in the individual pointing map for all of the sources. For the ALMA data, we generated boxes  {with a box size of about $50\times50$ pixels} centered on each source in the mosaics. In both cases, we fitted both a Gaussian and a background level, {with the background being treated as a free parameter} (\texttt{dooff} = True) since there was still extended emission around the sources. 

Table \ref{fitting_result} gives the source fitting results from \texttt{imfit}. For each source, we give the size and position angle (deconvolved from the beam) and total flux at each observed frequency. Errors are from the fit alone and do not include calibration uncertainties,  {which we estimate to be 10\%}.

\subsection{SED Fitting} \label{subsec:SED-fitting}

We used the NOEMA and ALMA multi-wavelength maps to construct SED profiles for the six protostellar cores along the OMC 2/3 filament.  Thermal dust emission can be described using the modified black body function \citep[e.g.,][]{mason2020}: 
\begin{equation}
  I_{\nu,dust}\propto \frac{\nu^{3+\beta}}{\exp(h\nu/kT_{d})-1}
  \label{MBB}
\end{equation}

\noindent where the dust opacity term has  been replaced with a power law given by $\kappa_\nu \propto \nu^\beta$. This model is generally seen as an excellent description of dust emission at millimeter and submillimeter wavelengths, with typical values of $1.5 < \beta < 2.5$ on filament and cloud scales \citep[e.g.,][]{Goldsmith1997, Sadavoy2013, Dib2020}. For low frequency emission (long wavelengths) on the Rayleigh-Jeans tail, $h\nu/kT_d \ll 1$ and Equation 1 simplifies to:

\begin{equation}
I_{\nu,dust} \propto \nu^{2+\beta} \equiv \nu^\alpha
\label{MBB2}
\end{equation}

\noindent where $\alpha = 2+ \beta$ is called the spectral index.  Since all our observations are at low frequencies of $\nu < 180$ GHz and  {we do not have robust constraints on the dust temperature for these targets on the observed scales}, we used Equation \ref{MBB2} to measure the spectral slope, $\alpha$ and infer $\beta$.  

We implemented a Monte Carlo Markov Chain (MCMC) code to measure $\alpha$ from the SED slope. To optimize the fitting process, we utilized the \texttt{emcee} code \citep{emcee2013} with an assumed, simple power-law model of $S_\nu = A\nu^{\alpha}$. To test how the slope changes for each source, we fitted power-law functions to the NOEMA data and ALMA-ACA data separately.  When running \texttt{emcee}, we employed uniform priors on the parameter $\alpha$, with bounds set as $-5.0 < \alpha < 5.0$. We also performed \texttt{emcee} with 32 walkers, enabling the walkers to rapidly explore the entire posterior distribution.   {Table \ref{table_SEDresult} gives the best fit $\alpha$ and $\beta$ slopes obtained from the simple power-law fitting applied to our data}. 

Since the flux errors reported in Table \ref{fitting_result} do not include flux calibration uncertainties, we further used a Monte Carlo approach to better determine the power-law slope error from the ALMA Band 4 and Band 5 data, which each have estimated flux calibration uncertainties of 10\%.  {However, it should be noted that the calibration error at 177 GHz may be larger due to its proximity to the 183 GHz atmospheric water line wing, potentially affecting the accuracy of flux measurements at this frequency. We used the Monte Carlo method to determine the uncertainty in the SED slopes from flux calibration errors because these errors are correlated for Band 4 and Band 5.} We used a random selector to generate two sets of 5000 samples from a Gaussian distribution with mean 1 and standard deviation of 0.1 to represent the 10\% flux calibration uncertainty in each band. We applied the randomly-selected values to the observed data and re-fitted the power law slopes. We then used the \nth{16} and \nth{84} percentiles of best-fit slopes to represent the error. For the NOEMA data, since the data were taken simultaneously with the same flux calibration, the flux calibration errors will not affect the slope.

\section{Results}
\begin{figure*}
 \centering
 \includegraphics[width=13.5cm]{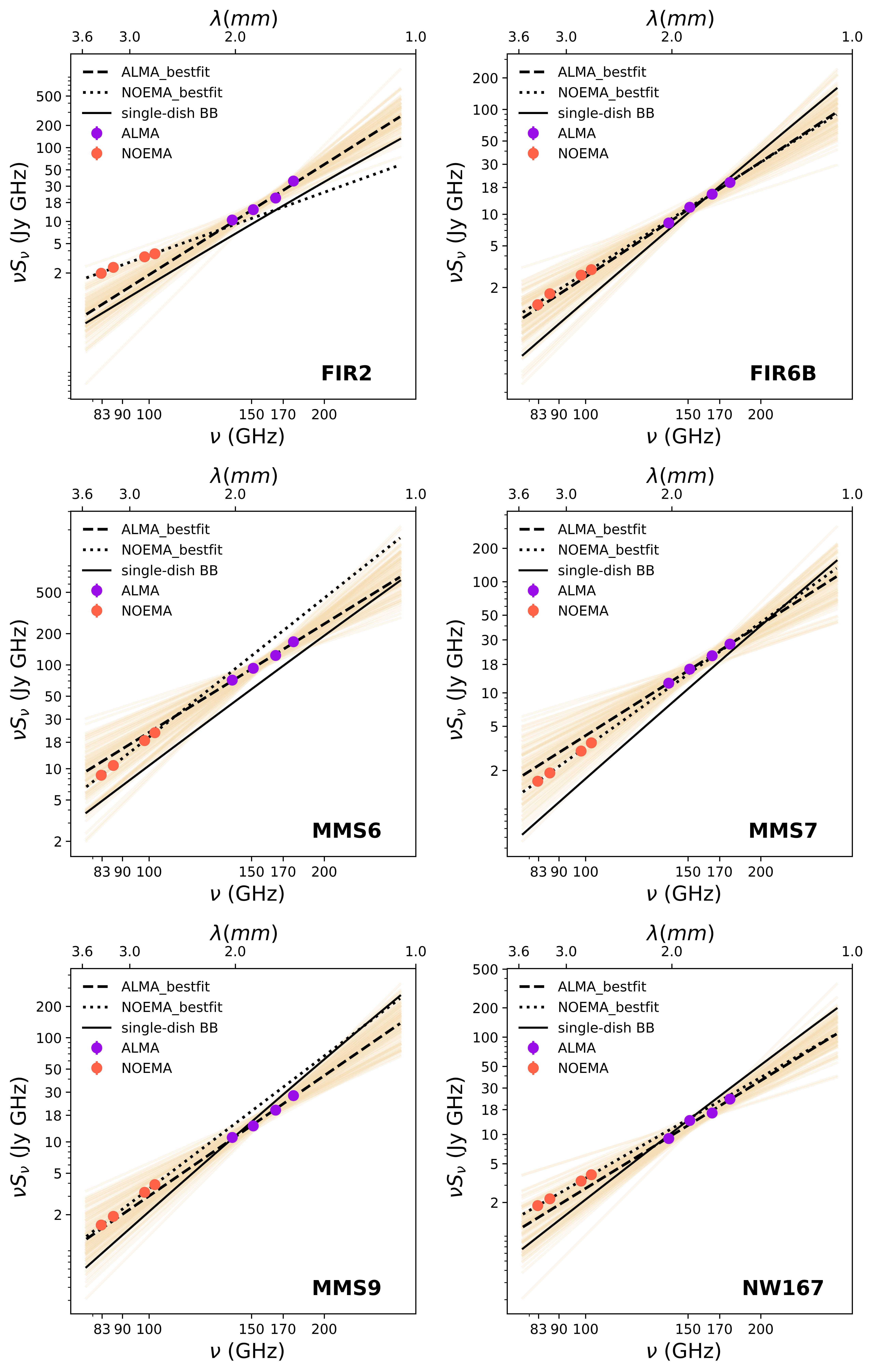}
 \caption{SEDs for the six sources with best-fit power-law slopes for the ALMA data (dashed) and NOEMA data (dotted). The yellow areas show the uncertainty on the ALMA slope from 10\% flux calibration errors. Solid lines show the modified blackbody functions from single-dish data (\emph{Herschel} at 0.16 -- 0.5 mm and IRAM+GISMO at 2 mm) from \citet{Sadavoy2016}.}
 \label{fig:omc_result}
\end{figure*}

Figure \ref{fig:omc_result} shows the SEDs of all sources using the NOEMA and ALMA data from 83 GHz to 177 GHz ($\sim 1.7 - 3.6$ mm). Plotted also are power-law best-fit slopes for the ALMA data alone (dashed lines for ALMA best fit and yellow area for the Monto Carlo uncertainties) the NOEMA data alone (dotted lines), and the best-fit modified black body function obtained using single-dish data ($\beta$-SD; solid lines) from \citet{Sadavoy2016} using data from \emph{Herschel} and IRAM (0.16 -- 2 mm).

Broadly, the NOEMA data spanning 82.7 -- 102.3 GHz (2.9 -- 3.6 mm) show elevated fluxes relative to the single-dish modified black body function (solid line), which is consistent with \citet{Sadavoy2016}.  It is interesting that the NOEMA data are still considerably elevated considering that these data are from interferomtery which filters out emission on scales $> 40$\arcsec, whereas the single-dish data, including the prior GBT data at 3 mm, are sensitive to emission on larger scales.  This excess flux at $\lambda \gtrsim 2.9$ mm suggests that the NOEMA data (and similarly the GBT data) are dominated by emission coming from a different origin than the emission captured by the single-dish data at 0.16 -- 2 mm.

The ALMA data yield more mixed results.  In some cases, the ALMA data are elevated over the single-dish modified blackbody function (solid line), whereas in other cases it is below or in agreement with the single-dish trend. As with the NOEMA observations, the ALMA data are filtered on scales $> 40$\arcsec, so we would expect that emission at comparable wavelengths (e.g., the single-dish data is  constrained by emission at 145 GHz (2 mm) at 22\arcsec\ resolution from the IRAM 30m telescope) to be less for the ALMA data.  Nevertheless, we caution against over-interpreting the difference in flux between ALMA and the modified blackbody fit from the single-dish data as differences in source extraction and map resolution could also be a factor.

\begin{longtable*}{ccccc}
\caption{Source Fitting Results}\label{fitting_result}\\

\hline
$\nu$ &$a$ & $b$ & PA & $S_{\nu}$ \\
(GHz) &(arcsec) & (arcsec)& (deg) & (mJy) \\
\hline
\endfirsthead

\multicolumn{5}{c}{{\bfseries \tablename\ \thetable{} -- continued from previous page}}\\
\hline
$\nu$ &$a$ & $b$ & PA & $S_{\nu}$ \\
(GHz) &(arcsec) & (arcsec)& (deg) & (mJy) \\
\hline
\endhead

\hline
\multicolumn{5}{c}{Continued on next page}\\
\hline

\endfoot
\hline
%\insertTableNotes

\multicolumn{5}{l}{$^a$ indicates that the source could not be deconvolved from the beam.} \\

\endlastfoot

\multicolumn{5}{c}{FIR2} \\
\hline

82.71897 &  {9.5 $\pm$ 1.4} &  {4.9 $\pm$ 0.8} &  {94 $\pm$ 10} &  {24.0 $\pm$ 1.3} \\

86.78157 &  {8.2 $\pm$ 1.2} &  {5.4 $\pm$ 0.6} &  {86 $\pm$ 17} &  {27.5 $\pm$ 1.4}\\

98.20925 &  {6.4 $\pm$ 0.8} &  {4.8 $\pm$ 2.0} &  {14 $\pm$ 47} &  {33.8 $\pm$ 2.0}\\

102.2718 &  {6.2 $\pm$ 1.0} &  {4.0 $\pm$ 3.1} &  {2.5 $\pm$ 48.4} &  {35.7 $\pm$ 2.2} \\

138.9607 &  {$< 13.4^a$} &  {$< 11.1^a$} & ... &  {75.2 $\pm$ 7.8}\\

150.9918 &  {$< 12.9^a$} &  {$< 10.8^a$} &  {...} &  {95.7 $\pm$ 9.7}\\

164.9749 &  {6.7$\pm$ 1.4} &  {1.5$\pm$3.2} &  {167$\pm$22} &  {125.9 $\pm$ 9.3}\\

177.007 &  {8.7 $\pm$ 1.3} &  {2.9 $\pm$ 2.4} &  {166 $\pm$ 14} &  {199 $\pm$ 15}\\

\hline

\multicolumn{5}{c}{FIR6B} \\

\hline

82.71897 &  {$< 14.4^a$} &  {$< 9.0^a$} & ... &  {16.6 $\pm$ 1.2} \\

86.78157 &  {3.4 $\pm$ 1.8} &  {1.0 $\pm$ 2.4} &  {36 $\pm$ 47} &  {20.1 $\pm$ 1.2} \\

98.20925 &  {3.2 $\pm$ 0.7} &  {1.4 $\pm$ 1.4} &  {15 $\pm$ 56} &  {26.7 $\pm$ 1.1} \\

102.2718 &  {3.0 $\pm$ 0.8} &  {1.8 $\pm$ 1.2} &  {3.9 $\pm$ 76.7} &  {29.0 $\pm$ 1.1}\\

138.9607 &  {$< 14.2^a$} &  {$< 9.2^a$} & ... &  {59.4 $\pm$ 3.1}\\

150.9918 &  {$< 12.9^a$} &  {$< 9.2^a$} & ... &  {77.1 $\pm$ 4.1}\\

164.9749 &  {4.4$\pm$0.8} &  {1.8$\pm$1.9} &  {4.3$\pm$45.5} &  {94.2 $\pm$ 4.6}\\

177.007 &  {7.7$\pm$1.0} &  {2.2$\pm$2.0} &  {31$\pm$62} &  {113.6 $\pm$ 6.6}\\

\hline

\multicolumn{5}{c}{MMS6} \\

\hline

82.71897 &  {$< 14.1^a$} &  {$< 8.2^a$} & ... &  {104.8 $\pm$ 3.6} \\

86.78157 &  {$< 12.9^a$} &  {$< 8.9^a$} & ... &  {123.9 $\pm$ 4.0}\\

98.20925 &  {$< 11.4^a$} &  {$< 8.4^a$} & ... &  {190.1 $\pm$ 5.0} \\

102.2718 &  {3.2 $\pm$ 0.4} &  {0.8 $\pm$ 1.4} &  {90.1 $\pm$ 1.7} &  {217.6 $\pm$ 5.3} \\

138.9607 &  {$< 13.8^a$} &  {$< 9.8^a$} & ... &  {514 $\pm$ 15} \\

150.9918 &  {$< 13.2^a$} &  {$< 9.3^a$} & ... &  {614 $\pm$ 13} \\

164.9749 &  {3.9 $\pm$ 0.6} &  {0.9 $\pm$ 1.7} &  {164 $\pm$ 21} &  {748 $\pm$ 25} \\

177.007 &  {4.2 $\pm$ 0.5} &  {1.8 $\pm$ 1.1} &  {163 $\pm$ 18} &  {945 $\pm$ 31} \\

\hline

\multicolumn{5}{c}{MMS7} \\

\hline

82.71897 &  {$< 14.2^a$} &  {$< 9.9^a$} & ... &  {19.4 $\pm$ 0.9}\\

86.78157 &  {$< 13.2^a$} &  {$< 9.4^a$} & ... &  {21.9 $\pm$ 1.0}\\

98.20925 &  {$< 11.5^a$} &  {$< 8.8^a$} & ... &  {30.5 $\pm$ 1.0}\\

102.2718 &  {$< 10.8^a$} &  {$< 8.6^a$} & ... &  {34.7 $\pm$ 1.1}\\

138.9607 &  {5.9 $\pm$ 1.7} &  {3.8 $\pm$ 2.3} &  {14 $\pm$ 83} &  {87.9 $\pm$ 6.6} \\

150.9918 &  {5.4 $\pm$ 1.6} &  {3.5 $\pm$ 2.2} &  {14 $\pm$ 80} &  {108.1 $\pm$ 7.8} \\
 
164.9749 &  {7.1$\pm$1.2} &  {2.1$\pm$2.5} &  {20$\pm$16} &  {130.8 $\pm$ 8.6} \\

177.007 &  {6.1 $\pm$ 1.4} &  {3.0 $\pm$ 2.1} &  {23 $\pm$ 28} &  {155 $\pm$ 12} \\

\hline

\multicolumn{5}{c}{MMS9} \\

\hline

82.71897 &  {$< 13.6^a$} &  {$< 9.5^a$} & ... &  {19.2 $\pm$ 1.2} \\

86.78157 &  {$< 12.9^a$} &  {$< 9.0^a$} & ... &  {22.2 $\pm$ 1.3}\\

98.20925 &  {$< 11.4^a$} &  {$< 8.6^a$} & ... &  {33.4 $\pm$ 1.4}\\

102.2718 &  {3.6 $\pm$ 0.8} &  {1.5 $\pm$ 1.6} &  {29$\pm$33} &  {38.0 $\pm$ 1.4} \\

138.9607 &  {5.3 $\pm$ 1.8} &  {3.3 $\pm$ 2.0} &  {33 $\pm$ 48} &  {79.8 $\pm$ 5.3} \\

150.9918 &  {3.8 $\pm$ 1.7} &  {2.4 $\pm$ 1.8} &  {14 $\pm$ 86} &  {94.3 $\pm$ 6.0} \\

164.9749 &  {5.7 $\pm$ 1.3} &  {2.7 $\pm$ 2.2} &  {24 $\pm$ 34} &  {122.4 $\pm$ 8.6} \\

177.007 &  {5.5 $\pm$ 1.3} &  {2.5 $\pm$ 1.9} &  {28 $\pm$ 27} &  {157 $\pm$ 10} \\

\hline

\multicolumn{5}{c}{NW167} \\

\hline 

82.71897 &  {7.1 $\pm$ 0.6} &  {3.9 $\pm$ 2.5} &  {5.8 $\pm$ 17.2} &  {22.4 $\pm$ 0.8}\\

86.78157 &  {6.4 $\pm$ 0.5} &  {3.3 $\pm$ 2.1} &  {5.4 $\pm$ 14.2} &  {25.2 $\pm$ 0.8} \\

98.20925 &  {5.9 $\pm$ 0.5} &  {3.7 $\pm$ 1.1} &  {18 $\pm$ 14} &  {33.8 $\pm$ 0.9} \\

102.2718 &  {5.5 $\pm$ 0.4} &  {3.8 $\pm$ 0.8} &  {11 $\pm$ 17} &  {37.7 $\pm$ 1.0} \\

138.9607 &  {$< 13.9^a$} &  {$< 11.4^a$} & ... &  {65.4 $\pm$ 2.5} \\

150.9918 &  {7.9 $\pm$ 0.8} &  {4.0 $\pm$ 2.9} &  {12 $\pm$ 12} &  {92.1 $\pm$ 3.9} \\

164.9749 &  {7.3 $\pm$ 0.8} &  {3.3 $\pm$ 1.5} &  {3.4 $\pm$ 10.8} &  {100.8 $\pm$ 4.5} \\

177.007 &  {7.3 $\pm$ 0.8} &  {4.5 $\pm$ 1.6} &  {0.0 $\pm$ 15.0} &  {130.5 $\pm$ 6.1} \\

\end{longtable*}

Hereafter, we focus on the slopes instead of the absolute flux values in our analysis as relative values (slopes) should be more robust against uncertainties from flux calibration, flux extraction, and spatial filtering.  Table \ref{table_SEDresult} lists the spectral indices ($\alpha$) and subsequent dust opacity indices ($\beta$) for each target as measured from the NOEMA and ALMA data alone. We fit separate slopes for the ALMA and NOEMA data due to evidence of a break (flattening) in the power-law profile at $\lambda \gtrsim 3$ mm.

Looking at the sample as a whole, we find that (1) the NOEMA and ALMA slopes tend to be shallower than the single-dish modified blackbody fits and (2) the NOEMA and ALMA slopes themselves tend to agree within $1 \sigma$ errors of each other.  The median $\beta$ indices are 0.74, 0.68, and 1.8 for the NOEMA, ALMA, and single-dish data, respectively, where the NOEMA and ALMA data have $\beta < 1.3$ in most cases and the single-dish data have $\beta \gtrsim 1.3$.  The exceptions to both of these trends are with FIR2 and MMS6, where the NOEMA and ALMA slopes differ by 2$\sigma$ and we measure a steeper slope for either NOEMA or ALMA compared to the single-dish fits. These distinctions suggest that FIR2 and MMS6 may be unusual sources among the entire sample.  We discuss the individual sources in more detail in the next section.

We note that our analysis used Equation \ref{MBB2} (power-law fits to the SEDs) rather than modified blackbody fit (e.g., Equation \ref{MBB}). For emission on the Rayleigh-Jeans tail, these two approaches should be equivalent. To verify this assumption, we calculated $\alpha$ from Equation \ref{MBB} assuming a typical dust temperature of 25 K \citep[e.g.,][]{Sadavoy2016, Friesen2017} and fixed $\beta = 1$ to represent the typical values in OMC 2/3. The inferred $\beta$ values derived from the $\beta = \alpha - 2$ assumption compared to the more accurate modified blackbody fit were approximately 15\% and 10\%  underestimated for the ALMA and NOEMA data, respectively. These differences are less than the errors reported in Table \ref{table_SEDresult} and therefore our assumption does not greatly affect our results. We nevertheless prefer to use a power-law fit over a modified blackbody fit because the temperatures themselves are unknown at these scales. Since we are targeting dust emission on $\sim$ 0.02 -- 0.08 pc ($\sim$ 4000 -- 16000 au) scales around luminous protostars, the dust temperatures could be much higher than the assumed 25 K (which was taken from single dish data that encompass larger scales). For these reasons, the spectral indices will better match the true modified blackbody slopes over assuming an incorrect temperature.

\begin{deluxetable}{cccccc}
\tablecaption{Best-fit spectral index slopes for NOEMA and ALMA\label{table_SEDresult}}
\tablewidth{0pt}
\tablehead{
\colhead{Source} & \colhead{$\alpha$-NOEMA} & \colhead{$\alpha$-ALMA} &
\colhead{$\beta$-NOEMA} & \colhead{$\beta$-ALMA} & \colhead{$\beta$-SD} 
}
\decimalcolnumbers
\startdata
FIR2 &  {1.84 $\pm$ 0.56} &  {3.95 $\pm$ 0.52} &  {-0.16 $\pm$ 0.56} &  {1.95 $\pm$ 0.52} & 1.3$\pm$0.1\\
FIR6B &  {2.51 $\pm$ 0.55} &  {2.64 $\pm$ 0.29} &  {0.51$\pm$ 0.55} &  {0.64 $\pm$ 0.29} & 1.9$\pm$0.1 \\
MMS6 &  {3.45 $\pm$ 0.30} &  {2.47 $\pm$ 0.17} &  {1.45 $\pm$ 0.30} &  {0.47 $\pm$ 0.17} &1.3$\pm$0.1\\
MMS7 &  {2.74 $\pm$ 0.39} &  {2.33 $\pm$ 0.42} &  {0.74 $\pm$ 0.39} &  {0.33 $\pm$ 0.42} &1.7$\pm$0.1 \\
MMS9 &  {3.24 $\pm$ 0.49} &  {2.84 $\pm$ 0.38} &  {1.24 $\pm$ 0.49} &  {0.84 $\pm$ 0.38} & 2.1$\pm$0.1 \\
NW167 &  {2.44 $\pm$ 0.30} &  {2.68 $\pm$ 0.23} &  {0.44 $\pm$ 0.30}  &  {0.68 $\pm$ 0.23} & 1.8$\pm$0.1 \\
\enddata
\tablecomments{The relationship between $\alpha$ and $\beta$ is defined as $\alpha = \beta + 2$. $\beta$-SD is measured using a modified black-body fit to single-dish data only from \citet{Sadavoy2016}.  {The observation frequencies are 81-104 GHz (NOEMA), and 137 -- 153 GHz and 163 -- 179 GHz for ALMA Band 4 and 5, respectively.}}
\end{deluxetable}

\section{Individual Source Results}
Below we discuss the results for each source. See Figure \ref{fig:omc_result} and Table \ref{table_SEDresult} for the SEDs and best-fit slopes.

\subsection{FIR 2}

FIR2 is a deeply embedded, low-mass protostar driving an outflow \citep{Takahashi2008,Tanabe2019}. It has a moderate luminosity of approximately $1.3 \, L_{\odot}$ and is surrounded by a flattened, relatively dense, infalling envelope with an estimated mass infall rate of $7.6 \times 10^{-6} \, M_{\odot} \, \text{yr}^{-1}$ \citep{Poteet2011}. FIR2 is the first protostar to show unambiguous evidence of crystalline silicates in absorption, likely due to thermal processing, with temperatures reaching 1000 K \citep{Poteet2011}, and  
significant free-free emission \citep{Bouvier2021}. \citet{Bouvier2021} found that $>50\%$ of the flux at 32.9 GHz could be attributed to free-free emission.

The SED for FIR2 is unusual among the other sources. It has the flattest slope from the NOEMA data and the only source with $\alpha < 2$ ($\beta < 0$).  Such a shallow slope at NOEMA frequencies is consistent with a high fraction of free-free emission \citep{Bouvier2021} (see Section \ref{subsec:FreeFree} for further discussion).  FIR2 also has the steepest slope measured from the ALMA data.   The slopes from the NOEMA and ALMA measurements disagree at a 2$\sigma$ level, which suggests that the ALMA and NOEMA data are tracing different material over the same scales.  Instead, the ALMA slope is more consistent with and even steeper than  the modified blackbody fit from the single-dish ($\beta$-SD) data. A steeper slope from the ALMA data over the single-dish data is unexpected as the ALMA data are more likely to trace emission from larger dust grains or warmer dust due to filtering out emission on scales $\gtrsim 0.08$ pc.  Both larger dust grains and strong temperature gradients should result in flatter SED profiles \citep[e.g., the $\beta$-temperature anti-correlation][]{Shetty2009}. A more careful and consistent comparison between the single-dish and ALMA data will be necessary to determine whether or not the difference in slope is significant.

\subsection{FIR6B}

FIR6B is a Class 0 intermediate-mass protostar with a bolometric luminosity of $L_{\text{bol}} = 21.9 \, L_{\odot}$ \citep{Furlan2016}. It is associated with fast-rotating protostellar jets, exhibiting a rotation velocity greater than $20 \, \text{km} \, \text{s}^{-1}$ and a specific angular momentum of about $10^{22} \, \text{cm}^2 \, \text{s}^{-1}$, 
a relative large value compared to most protostars \citep{Mezger1990,Chini1997,Furlan2016}. This rapid rotation is explained by a magnetohydrodynamic (MHD) disk wind, where magnetic fields efficiently transfer angular momentum from the disk to the jet \citep{BlandfordPayne1982}. High-resolution ALMA CO ($J=2\text{-}1$) observations reveal a velocity gradient along the short axis of the jet, with the launching radius estimated between $2.18$ and $2.96 \, \text{au}$, further supporting the disk wind hypothesis \citep{Matsushita2021}. Additionally, FIR6B drives a low-velocity bipolar outflow ($\sim 5 \, \text{km} \, \text{s}^{-1}$), extending from a blueshifted lobe to the northeast and a redshifted lobe to the southwest \citep{Takahashi2008,Shimajiri2009}. Given its powerful jet and detections at 3.6 cm from the VLA \citep{Reipurth1999}, we would expect FIR6B would have free-free emission, although we could find no detailed analysis to date.

FIR6B shows relatively consistent slopes for both the ALMA and NOEMA datasets.  The two slopes are consistent within 1$\sigma$, suggesting that they are tracing the same type of emission. Both slopes are also considerably more shallow than the single-dish modified blackbody fit.  The $\beta$ indices differ by $> 2\sigma$ and $> 3\sigma$ between the single-dish fits and the NOEMA and ALMA data, respectively. The higher discrepancy with the ALMA data is interesting as the single-dish fits include a data point at 145 GHz from IRAM+GISMO at 22\arcsec\ (0.04 pc) resolution.   
Thus, it appears that the interferometry data are tracing a different continuum component than the single-dish data, such that the SED for this source may be dependent on scale.

\subsection{MMS6}

MMS6 is an extremely young intermediate-mass protostellar core with a mass of approximately \(0.29 \, M_{\odot}\) and a bolometric luminosity of $\lesssim$ \(60 \, L_{\odot}\) \citep{Takahashi2008}.  
MMS6 has a notably compact outflow, with a lobe size of approximately \(1500 \, \text{au}\), making it an unusually small molecular outflow  intermediate-mass protostars, suggesting that this source is at an early evolutionary stage \citep{Aso2000, Takahashi2008}. 
Furthermore, the core has a high density, estimated at \(1.5 \times 10^{10} \, \text{cm}^{-3}\), suggests it may be optically thick at \(850 \, \mu\text{m}\), reinforcing its classification as an intermediate-mass protostellar core. \citet{Reipurth1999} detected bright 3.6 cm emission toward MM6, which they attributed to free-free emission from ionized material.  
In contrast, \citet{Takahashi2009} measured a dust opacity index $\beta$ of approximately 0.93 using high resolution (160 au) multi-wavelength interferometric data between 0.9 mm and 7.3 mm, and \citet{Tobin2020} similarly find a spectral index of $\alpha = 2.7$ ($\beta = 0.7$) at higher resolutions (40 au) using ALMA and VLA data between 0.87 mm and 9 mm.  These measurements primarily trace the disks and inner envelope of MMS6, which is on much smaller scales that what we observe ( 4000 -- 16000 au) and the value of $\beta$ measured at such long wavelengths support that the emission is primarily coming from dust instead of free-free emission.

Using the ALMA and NOEMA data, we measure spectral indices that do not agree with the previous measurements.  For the ALMA data, we get a shallower slope of $\beta = 0.47$ and for the NOEMA data, we get a steeper slope of $\beta = 1.45$.  This suggests that the previous work of \citet{Takahashi2009} and \citet{Tobin2020} may have measured an average of these slopes.  The SED produced by \citet{Takahashi2009} (see their Figure 5) does suggest a change in slope (flattening) between 2.3 mm and 3.3 mm, which our SED of MMS6 confirms.

Overall, MMS6 also appears to have different properties from the other objects in the sample.  All sources except MMS6 have systematically flatter slopes at the NOEMA frequencies compared to the single-dish data by $\gtrsim 2\sigma$. MMS6 presents an anomalous case, where we measure an identical (within $1\sigma$) slope between the NOEMA data and the single-dish data, although the NOEMA fluxes are still elevated above the single-dish extrapolations by roughly a factor of two. The NOEMA and ALMA slopes for MMS6 also disagree at $\sim 2\sigma$, suggesting that they are tracing different types of emission.  This result contradicts the previous conclusions that the data between 2.3 mm and 7.3 mm is coming from dust emission \citep{Takahashi2009}, unless the dust properties themselves are changing (see Section \ref{subsec:dust emission}).   Instead, the flatter slope from the ALMA data transitioning to a steeper slope with the NOEMA data could mean that the NOEMA data are tracing optically thick free-free emission, which have a steeper spectral index than optically thin free-free emission (see Section \ref{subsec:FreeFree}).  Indeed, no significant free-free emission was detected at centimeter wavelengths \citep{Takahashi2009}, which supports optically thick free-free emission at $\lambda > 3 mm$.  Nevertheless, the flatter slope at the ALMA frequencies indicates that MMS6 must transition to optically thin free-free emission over a very narrow range of wavelengths between 2.9 mm and 2.15 mm otherwise we would not measure such a flat profile between 2.15 mm and 1.7 mm.  Therefore, we cannot make strong conclusions about free-free emission in MMS6.

\subsection{MMS7}

MMS7 is characterized by a single intermediate-mass protostar \citep{Takahashi2008b} with a bolometric luminosity of approximately \(42.9 \, L_{\odot}\) \citep{Furlan2016}. Initially identified as a Class 0 source \citep{Chini1997}, MMS7 was later reclassified as Class I \citep[e.g.,][]{Tobin2020}.  Recent ALMA observations have resolved MMS7 into two distinct components, HOPS-84-A and HOPS-84-B, both of which exhibit compact disk-like structures, with sizes of approximately \(95 \times 42 \, \text{au}\) and \(37 \times 17 \, \text{au}\), respectively \citep{Feddersen2020}.  
MMS7 also has a giant bipolar molecular outflow, with its blueshifted component displaying a cavity-like structure \citep{Aso2000}. It is also associated with free-free emission likely arising from radio jets based on VLA observations towards it \citep{Reipurth1999}.

The ALMA and NOEMA data yield slopes that are in agreement within 1$\sigma$ of each other, which implies that the emission characteristics are similar across different instruments and frequencies. Nevertheless, both measurements have relatively large errors, which suggest that the power-law slopes for MMS7 are not well constrained.  When comparing ALMA and NOEMA data to single-dish results, MMS7’s slope appears to be shallower by $\gtrsim$ 2$\sigma$, reinforcing the idea that the higher resolution of ALMA data is tracing emission components that the single-dish data, with its lower resolution, might have smoothed out. This indicates that the slope variations in MMS7 are likely dependent on the spatial scales probed by the different observations.

\subsection{MMS9}

MMS9 is a well-known intermediate-mass protostar characterized by several outflowing sources indicating multiple star formation activity \citep{Chini1997}.  It has a (combined) bolometric luminosity of \(8.9\, L_{\odot}\) and bolometric temperature of 38 K \citep{Tobin2020}.  
Due to the multiple outflows and jets, it is not surprising that MMS9 has evidence of substantial free-free emission \citep{Bouvier2021} (see Section \ref{subsec:FreeFree} for more details).

Like MMS7 and FIR6B, MMS9 shows consistent slopes across both ALMA and NOEMA data.  The two datasets yield spectral indices that agree within 1$\sigma$, indicating that they are likely tracing the same emission features.  Both slopes are also noticeably shallower than the modified blackbody fit from the single-dish data, but only the ALMA slope is significantly more shallow ($> 2\sigma$).  As with the previous cases, the discrepancy between ALMA and single-dish slopes may indicate a difference in dust emission with spatial resolution, with ALMA likely capturing finer-scale emission features. This suggests that the single-dish data are tracing more of the larger-scale emission from the surrounding environment, while ALMA provides insight into more compact, localized emission within MMS9.  We note that the multiple disks within MMS9 are unresolved by both the ALMA and NOEMA observations, however.

\subsection{NW167}

NW167 is a dense protostellar core identified in \citet{Chini1997}. NW167 is characterized by a bolometric luminosity of 6.2 $L_\odot$ and a bolometric temperature of 35.6 K, marking it as a Class 0 protostar in an early evolutionary stage with one detectable disk \citep{Tobin2020}. While more isolated than the other sources in the sample, NW167 is still part of a dense filamentary structure \citep{Bally1987,Peterson2008}.

NW167 exhibits a consistent (within 1 $\sigma$) slope between the ALMA and NOEMA data, suggesting that both datasets are tracing similar emission components. The single-dish data, however, has a significantly steeper slope by $> 3\sigma$ compared to both the ALMA and NOEMA observations. This difference is likely due to the difference in spatial scale between the interferometers and the single-dish observations, where the former focuses on the compact emission and the latter captures more extended emission.

\section{Discussion}\label{sec:Discussion}

Our findings suggests that there are some similarities and some differences between the target sources in terms of their SED properties.  Most of the sources have comparable (within $1\sigma$) slopes between the ALMA and NOEMA data.  Since we are tracing emission over comparable scales for both telescopes, consistent slopes likely indicate a common origin for the emission from 1.7 -- 3.6 mm.  Discrepancies with the single-dish modified blackbody fits are therefore likely due to differences in spatial resolution.  Nevertheless, some sources showed significant differences between the ALMA and NOEMA slopes, which suggest that different mechanisms may be producing the continuum data on these scales.  These objects are the best candidates for contamination (e.g., from free-free emission, AME) as such emission will likely affect the lower frequency NOEMA data more than the higher frequency ALMA data.

In the following sections, we investigate potential explanations for the changing SED slopes, encompassing various factors, namely, free-free emission, AME, and disk contamination. Additionally, we delve into the consideration of complex dust grain opacity functions. These considerations aim to provide a comprehensive understanding of the underlying processes that can contribute to the flattened SED profiles and the associated variations in fluxes at longer wavelengths of protostellar cores within the OMC 2/3 filamentary region.

\subsection{Spectral Line Contamination} \label{subsec:line cont}

Broadband continuum detectors can measure elevated fluxes due to transitions of the brightest molecular species in the bandpass \citep[e.g.,][]{Drabek2012,Schnee2014}. Both NOEMA and ALMA, however, allow us to identify and exclude channels likely contaminated by strong line emission. Since we excluded channels that exhibited strong line features, the flattened slopes are not caused by strong lines in the bandpass. 

\begin{figure*}
 \centering
 \includegraphics[trim = 3mm 0mm 7mm 0mm,clip=true,width=0.98\textwidth,height=12cm]{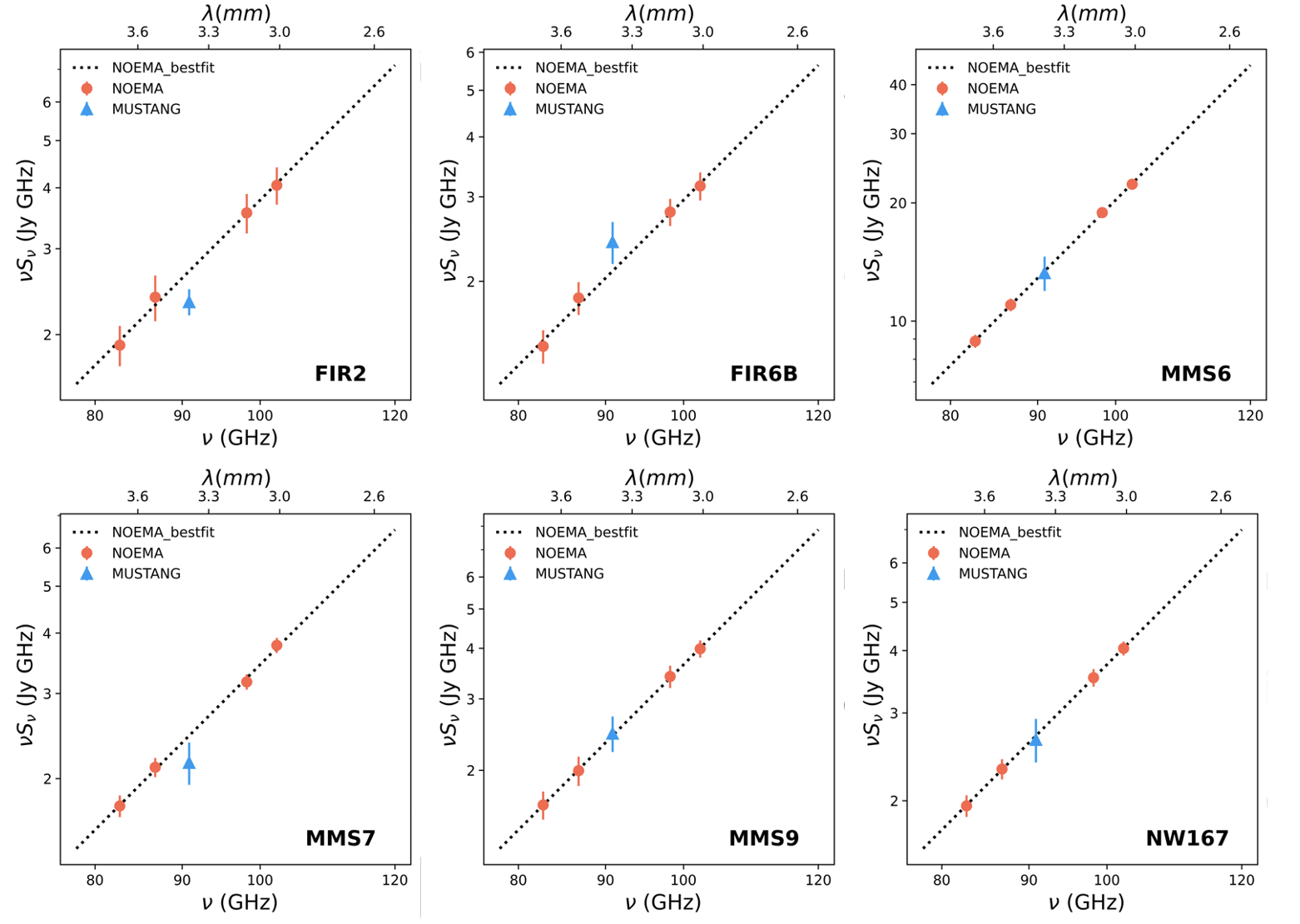}
 \caption{SEDs for the six sources with best-fit power-law slopes for the NOEMA data (dotted). The 3 mm GBT MUSTANG \citep{mason2020} are included for comparison but were not used in the fitting.  {To facilitate a robust comparison, the NOEMA data were smoothed to match the 10.8 \arcsec\ resolution of the GBT data.}}
 \label{fig:NOEMA_MUSTANG_result}
\end{figure*}

Moreover, we find excellent agreement between the NOEMA data and the 90 GHz (3.3 mm) data obtained from MUSTANG presented by \cite{Schnee2014}. Figure \ref{fig:NOEMA_MUSTANG_result} compares the GBT observations with the NOEMA observations for each target.  For consistency, we smoothed the NOEMA data to a common resolution of 10.8\arcsec\ to match the resolution of the GBT data \citep{Schnee2014}.  We find that the GBT measurements are within $30\%$ of the SED outlined by the 10.8\arcsec-smoothed NOEMA data.  This agreement lends further support to the conclusion that spectral line contamination was not a contributing factor to the earlier analyses of OMC 2/3 \citep[e.g.,][]{Schnee2014,Sadavoy2016,mason2020}.  Instead, this agreement suggests that even though the GBT is a single-dish data that recovers emission on larger scales than what is traced by NOEMA, the two facilities are mostly capturing emission from small scales.  This result is not unsurprising as longer wavelength emission primarily traces environments with higher densities and larger dust, which are mostly on smaller scales.

\subsection{Free-Free Emission} \label{subsec:FreeFree}

Thermal bremsstrahlung, commonly known as ``free-free'' emission, involves the emission of photons by accelerated charged particles and occupies a frequency range between synchrotron and thermal dust emission, typically spanning from 10 GHz to 100 GHz \citep[e.g.,][]{Draine2011}. Extended thermal free–free is more dominant in ionized regions of high-mass star formation like OMC 1 \citep[e.g.,][]{Dicker2009}, whereas OMC 2/3 lacks massive stars or star clusters \citep[e.g.,][]{Nielbock2003}. Previous estimates by \citet{mason2020} using 8.3 GHz VLA maps from \citet{Reipurth1999} found that both optically thin and optically thick free-free emission at $\sim$100 GHz is likely negligible based on flux density ratios.

In a more comprehensive study of free-free emission in OMC 2/3, \citet{Bouvier2021} used data from ALMA at 333 and 246.2 GHz and VLA at 32.9 GHz to determine the fraction of 32.9 GHz emission that could arise from free-free emission. They found larger values for $\alpha$, by factors of 3 -- 4, when analyzing data between 333 -- 246.2 GHz compared to the indices between 246.2 -- 32.9 GHz and 333 -- 32.9 GHz, suggesting the presence of free-free emission at 32.9 GHz. \citet{Bouvier2021} only quantified the portion of free-free emission in two of our targets, FIR2 and MMS9, however. Table \ref{FF} displays their estimated free-free emission fractions at 32.9 GHz based on the analysis of \citet{Bouvier2021}. Specifically, Table \ref{FF} compares the $\beta$ values obtained here from our NOEMA data (82.7 -- 102.3 GHz) with the corresponding measured $\beta$ index from \citet{Bouvier2021} using pairs of frequencies between 32.9 GHz and 333 GHz and a set of assumed temperatures.  The final column in Table \ref{FF} gives the fraction of 32.9 GHz emission that could arise from free-free emission.

From Table \ref{FF}, we can see that for sources like FIR2 the derived free-free emission at 32.9 GHz ranges between 55\% and 100\%, and some of the embedded protostars in MMS9 may also contribute substantial free-free emission. In both cases, we could be looking at most if not all flux (cumulative) at 32.9 GHz being due to free-free emission. This indicates that free-free emission is present in these sources. Nevertheless, our data are at $>$ 82 GHz, however, and dust emission will be more significant at 82 GHz than at 33 GHz. So while free-free emission may be prominent at 33 GHz, if the emission becomes optically thin, it will not be a significant factor at 86 GHz.  Given that we measure a spectral index of $\alpha \approx 3$ for MMS9 from the NOEMA data, we can expect that dust emission is present at the NOEMA frequencies for this source.  FIR2, however, has as NOEMA spectral index of $\alpha \approx 2$, which is consistent for optically thick free-free emission \citep{mason2020}.  Nevertheless, the free-free emission would have to transition to optically thin very quickly beyond the NOEMA frequencies otherwise we would detect a similar slope in the ALMA bands.  But for FIR2, the ALMA slope is much steeper ($\alpha \approx 4$) and more consistent with traditional dust than dust emission contaminated with free-free emission.

\begin{deluxetable}{cccc} 
%\tablenum{5}
\tablecaption{Free-Free emission for FIR2 and MMS9}\label{FF}
\tablewidth{0pt}
\tablehead{
\colhead{Source} & \colhead{$\beta$-NOEMA} &\colhead{$\beta$-B21} & \colhead{FF [\%]}  
}

\startdata
FIR2 & -0.07 $\pm^{0.60}_{0.61}$ &0.4 - 1.6 & 55 - 100 \\
MMS9-a & 0.98 $\pm$ 0.43 &0.3 - 0.4 & 0 - 23 \\
MMS9-b & $\cdots$ &0 & 0 - 22 \\
MMS9-c & $\cdots$ &0.3 - 1.8 & 47 - 100\\ 
MMS9-d & $\cdots$ &0 - 2.0 & $\cdots$ \\ 
\enddata
\tablecomments{The $\beta$-21 values are from \citet{Bouvier2021} and represent the range measured from data at 333-246.2 GHz, 246.2-32.9 GHz, and 333-32.9 GHz for a set of assumed temperatures between 10 - 200 K. The last column indicates the possible free-free contamination at 32.9 GHz, not at the NOEMA/ALMA frequencies.}
\end{deluxetable}

\subsection{Anomalous Microwave Emission (AME)} \label{subsec:AME}

The presence of AME has emerged as a potential explanation for the flattened slope observed in the SED at $<150$ GHz ($>2$ mm) \citep[e.g.,][]{kogut1996,Leitch1997,Oliveira-Costa1997}. It is predominantly attributed to electric dipole radiation emanating from small, charged spinning dust grains \citep[e.g.,][]{Draine1998, Hoang2016}. While alternative models involving magnetic dipole emission from large ferrous dust grains or spinning nano-silicates have been proposed, the most compelling explanation for AME lies in electric dipole emission from ultrasmall grains \citep[e.g.,][]{Draine1999,Hoang2016,Greaves2018}.

\citet{mason2020} examined the possibility of AME affecting the observed SEDs within OMC 2/3. Its typical peak frequencies usually fall within the 20-60 GHz range, where the peak can vary depending on local physical conditions, including grain size distribution, radiation intensity, dipole moment of the emitting grains, number density of grains, and ionization fractions of hydrogen and carbon \citep[e.g.,][]{Finkbeiner2004,Planck2011,Planck2015}. To explore AME in different environments, \citet{mason2020} utilized the SPDUST2 \citep[e.g.][]{Ali2009,Silsbee2011} code to derive representative AME spectra, encompassing both diffuse molecular clouds and dark clouds. Upon comparing these AME models with actual observed data from the OMC 2/3 region, however, the models didn't match the observed data.

Our SED slopes over the NOEMA frequencies support the idea that AME is not a major factor.  We see no evidence of a turnover in the spectral indices such that flux increases with decreasing frequency (e.g., $S_{\nu} \propto \nu^\alpha$ where $\alpha < 0$), which is expected for AME which peaks at frequencies below the observed NOEMA frequencies of 80 -- 100 GHz.  While we cannot rule out smaller contributions from AME without further investigation and more detailed modeling, we can conclude that AME is not a dominant feature at these scales.

\begin{deluxetable*}{ccccccccccccc}
\tablecaption{Fraction of NOEMA and ALMA emission to arise from protostellar disks}\label{diskinfo}
\tablewidth{0pt}
\tablehead{
    \colhead{Frequency (GHz)} & 
    \multicolumn{2}{c}{FIR2} & 
    \multicolumn{2}{c}{FIR6B} & 
    \multicolumn{2}{c}{MMS6} & 
    \multicolumn{2}{c}{MMS7} & 
    \multicolumn{2}{c}{MMS9} & 
    \multicolumn{2}{c}{NW167} \\[-2mm]
    \colhead{} & \colhead{$\alpha = 2$} & \colhead{$\alpha = 3$} &
    \colhead{$\alpha = 2$} & \colhead{$\alpha = 3$} & 
    \colhead{$\alpha = 2$} & \colhead{$\alpha = 3$} & 
    \colhead{$\alpha = 2$} & \colhead{$\alpha = 3$} & 
    \colhead{$\alpha = 2$} & \colhead{$\alpha = 3$} & 
    \colhead{$\alpha = 2$} & \colhead{$\alpha = 3$}\\[-8mm]
}
\decimalcolnumbers
\startdata
$S_{disk}$ (Jy) & \multicolumn{2}{c}{0.088} &  \multicolumn{2}{c}{0.34} &   \multicolumn{2}{c}{1.59} &   \multicolumn{2}{c}{0.46} &   \multicolumn{2}{c}{0.34} &   \multicolumn{2}{c}{0.21}  \\ 
82.71897 &  {22\%} &  {5\%} &  {100\%} &  {30\%} &  {91\%} &  {22\%} &  {100\%} &  {35\%} &  {100\%} &  {26\%} &  {56\%} &  {14\%} \\
86.78157 &  {21\%} &  {5\%} &  {100\%} &  {29\%} &  {84\%} &  {22\%} &  {100\%} &  {35\%} &  {100\%} &  {26\%} &  {55\%} &  {14\%} \\
98.20925 &  {22\%} &  {6\%} &  {100\%} &  {31\%} &  {70\%} &  {20\%} &  {100\%} &  {37\%} &  {86\%} &  {25\%} &  {52\%} &  {15\%} \\
102.2718 &  {23\%} &  {7\%} &  {100\%} &  {32\%} &  {67\%} &  {20\%} &  {100\%} &  {37\%} &  {82\%} &  {25\%} &  {51\%} &  {15\%} \\ 
138.9607 &  {20\%} &  {8\%} &  {96\%} &  {40\%} &  {52\%} &  {21\%} &  {88\%} &  {36\%} &  {72\%} &  {29\%} &  {54\%} &  {22\%} \\
150.9918 &  {18\%} &  {8\%} &  {88\%} &  {39\%} &  {52\%} &  {23\%} &  {85\%} &  {38\%} &  {72\%} &  {32\%} &  {45\%} &  {20\%} \\
164.9749 &  {17\%} &  {8\%} &  {86\%} &  {42\%} &  {50\%} &  {25\%} &  {84\%} &  {41\%} &  {66\%} &  {32\%} &  {49\%} &  {24\%} \\
177.007 &  {12\%} &  {6\%} &  {82\%} &  {43\%} &  {47\%} &  {25\%} &  {81\%} &  {42\%} &  {59\%} &  {31\%} &  {44\%} &  {23\%} \\ 
\enddata
\tablecomments{$S_{disk}$ flux values are measured at 0.87 mm (338.5 GHz) and represent the sum of all disks in the targets from \citet{Tobin2020}. Fractions greater than 100\% are represented by 100\%.}
\end{deluxetable*}

\subsection{Disk Contamination} \label{subsec:disk emission}

Protoplanetary disks around young protostars contain dust grains that can range in size from sub-micron to millimetre scales \citep[e.g.,][]{Villenave2020,Villenave2022, miotello2022}. Emission at millimetre wavelengths is sensitive to radiation from larger dust grains, which are expected to be biased to the mass distribution within the disk \citep[e.g.,][]{Tobin2020}. We determined the contribution of dust emission from the embedded protostellar disks within each source to determine whether or not these could be biasing the NOEMA and ALMA fluxes and ascertain whether disk dust emission can explain the elevated flux densities detected in some sources.

For simplicity, we extrapolated ALMA observations at 340 GHz (0.87 mm) with 40 au resolution from \citet{Tobin2020}  using a power-law relationship, $S_{\nu} \sim \nu^\alpha$ to infer the contribution of disk emission.  We then calculated the fraction of the observed NOEMA and ALMA fluxes that could be attributed to disk emission following,

\begin{equation}
  Fraction=\frac{S_{disk,e}}{S_{obs}}= \frac{S_{disk} \times \left(\frac{\nu_e}{338.5\ \mathrm{GHz}}\right)^\alpha}{S_{obs}}
  \label{disk_fraction}
\end{equation}

\noindent where $S_{disk,e}$ is the extrapolated disk flux at an extrapolated frequency of $\nu_e$, $S_{disk}$ is the disk flux from \citet{Tobin2020} observed at 338.5 GHz\footnote{We adopted a frequency of 338.5 GHz, corresponding to the average of the center frequencies of the two continuum sidebands in \citet{Tobin2020}.}, $S_{obs}$ is our measured NOEMA and ALMA flux (See Table \ref{fitting_result}), and $\alpha$ is spectral index.  We adopted two different spectral indices, $\alpha = 2$ and $\alpha = 3$, corresponding to values of $\beta =$ 0 or 1, which are representative of typical values in disks \citep[e.g.,][]{Tychoniec2020,miotello2022}.  Therefore, values of $\alpha =2$ likely represents an upper limit on disk contamination and $\alpha = 3$ are a lower limit.

 {Although the 340 GHz ALMA observations from \citet{Tobin2020} filtered out dust emission in the core and envelope, our focus is to estimate contamination coming from the disk alone.} 
 
To quantify the contribution of disk emission to the total flux for each protostellar core, we summed the contributions of all disks within the beam area of the NOEMA observations for each source. For example, four disks were summed for MMS9, while two disks were summed for MMS7. Once we estimated the flux contributions from the disks, we compared these measurements to the observed fluxes for each target. Table \ref{diskinfo} presents the estimated fraction of observed emission that could be from disk contamination at each frequency band.  
%Additionally, we performed the same analysis for ALMA Band 4 and 5 data, and the results are shown in the Table \ref{diskinfo} as well.

Table \ref{diskinfo} gives the outcomes from our measurements of potential disk contamination.  These results suggest that most sources could have substantial contamination from disks, especially for low values of $\alpha$.  Such low values of $\alpha$ are expected for optically thick disks or very large dust grains \citep[e.g.,][]{Carrasco-Gonzalez2019,Tychoniec2020,miotello2022}. From \citet{Tobin2020}, however, most of the sources likely have spectral indices steeper than 2.  The only exception is MMS6, which is more likely consistent with $\alpha = 3$, in agreement with previous measurements of $\beta \approx 0.5-1$ from \citet{Takahashi2009}.  Therefore, we consider $\alpha = 3$ for MMS6 and a value between $\alpha = 2$  and $\alpha = 3$ for the remaining targets in the following analysis.

We find that only for FIR2 ($\alpha = 2 - 3$) and MMS6 ($\alpha =3$, disk contamination is likely minimal.  The range of contamination is between $5-23$\%\ for FIR2 and between $20-25$\% for MM6.  This level of contamination cannot fully explain elevated fluxes at the NOEMA frequencies for these sources, which are higher than expected over the modified blackbody functions by up to a factor of 2. It is interesting to note that these two sources are also the only two sources with significantly ($> 2\sigma$) different SED slopes between the ALMA and NOEMA data, indicating that the ALMA and NOEMA emission likely arise from different mechanisms.  If disk contamination is a non-negligible factor in the other sources, then it would suggest that MMS6 and FIR2 have different SED profiles due to other factors dominating their SEDs.  Both FIR2 and MMS6 have dense envelopes \citep{Takahashi2008, Tobin2020} and evidence for free-free emission \citep{Reipurth1999, Bouvier2021} which could be dominating over the embedded disks.

For the other sources, we cannot rule out contamination from the embedded disks.  In particular, FIR6B, MMS7, and MMS9 could have nearly all of their flux at $\lesssim 100$ GHz coming from disk emission and non-negligible fractions ($> 60$)\%\ at $138-177$ GHz for $\alpha = 2$.  With steeper indices, however, these fractions drop to $25-45$\%, which suggests  that the emission could still be non-negligible, but not dominant.  NW167 has less significant contamination, ranging from $44-56$\%\ for $\alpha = 2$ to $< 25$\%\ for $\alpha = 3$.  Since we expect the true disk contamination of these sources to be somewhere in between these calculations, we cautiously suggest that disk contamination is most likely for FIR6B and MMS7 given their ranges and their high disk masses from \citet{Tobin2020} (see also Table \ref{tab:list of sources}).  These sources also have consistent (within 1$\sigma$) slopes between the ALMA and NOEMA data, suggesting that the common mechanism to explain their dust emission may be disk emission.  MMS9 and NW167 are the other two sources with consistent ALMA and NOEMA slopes, but the possibility of disk contamination for them is less clear.  MMS9 has a relatively low estimated disk mass but a high possible contamination fraction.  NW167 has a relatively high disk mass, but a lower possible contamination fraction (see Table \ref{tab:list of sources} for disk masses). %Both could have some contamination from its multiple embedded disks, but it depends on whether or not the disks have $\alpha =3$ or $\alpha =2$.
We also caution that we do not see strong evidence of corresponding emission excesses at 140 -- 180 GHz for these sources, which would be expected if the ALMA/NOEMA fluxes are elevated over the single-dish fluxes due to biases from disk emission.

This analysis implies that the disks within some sources could possess  enough mass to exert contamination on much larger scales ($>0.02$ pc or $> 4000$ au) than the disks themselves.   
Such contamination could also impact the single-dish GBT data, which have a comparable resolution to the smoothed NOEMA data and show similar fluxes as the NOEMA data (see Figure \ref{fig:NOEMA_MUSTANG_result}).  This result indicates that care is needed when interpreting higher resolution, low frequency observations either from single-dish telescopes or interferometers for protostars with embedded disks.

\subsection{Unusual Dust Properties} \label{subsec:dust emission}

An additional explanation for the flattened slopes observed in the SED profiles could be attributed to the intrinsic properties of the dust grains themselves. The formation of millimeter and sub-millimeter sized dust grains in the dense Orion cores, can create a flat spectral index compared to normal ISM dust grains \citep[e.g.,][]{Miettinen2012,Schnee2014}. The large grains could account for the low $\beta$ values measured. As such, the elevated fluxes observed at $<150$ GHz may be explained by a change in the dust opacity slope, where the true distribution could follow a broken power-law with a flatter slope at $\nu < 150$ GHz ($\lambda > 2$ mm). The formation of such large dust grains in protostellar envelopes is possible due to rapid aggregation processes occurring before young stellar objects enter the Class II stage. Early grain growth is supported by evidence of micron-sized grains in dense cores and millimeter-sized grains in Class 0 and I objects as evidenced by flat SED profiles \citep[e.g.,][]{Pagani2011, Schnee2014, Miotello2014,Galametz2019}. Nevertheless, we caution that from our aforementioned disk contamination study, large dust grains seen on envelope-scales toward unresolved disks could be biased by the disks. Therefore, modeling disk and envelope emission using visibilities over images would be necessary to break this degeneracy \citep[e.g., following][]{Miotello2014, Galametz2019, Cacciapuoti2023}.

Alternatively, sophisticated models derived from laboratory measurements of amorphous dust grains have produced flattened spectral slopes at millimeter wavelengths \citep[e.g.,][]{Boudet2005,meny2007,Coupeaud2011,Paradis2011}. This trend primarily stems from two-level system fluctuations occurring within the dust grains. The model proposed by \citet{Paradis2011} was able to partially explain the elevated emission at 100 GHz from the GBT, but it under predicts the 30 GHz fluxes \citep[e.g.,][]{mason2020}. Further comprehensive studies are required to gain a deeper understanding of the complex behavior of dust grains at millimeter wavelengths.

\section{Conclusions}\label{sec:Conclusion}

This study investigated an unexpected flux increase at 90 GHz in the OMC 2/3 region first noted by \citet{Schnee2014}. We used multi-wavelength observations from NOEMA and ALMA to characterize the SED profile of six targets in the OMC 2/3 filamentary region to identify potential contributing factors to explain the change in SED slope at $< 100$ GHz.

Our main conclusions are:

\begin{enumerate}

    \item NOEMA slopes confirm the flatter profiles at frequencies $< 150$ GHz (wavelengths $>2$ mm).   All sources except MMS6 give flatter SED slopes between 82.7 GHz and 102.3 GHz (2.9 -- 3.7 mm) compared to modified blackbody fits using single dish data (from 0.16 -- 2.0 mm).

    \item  The ALMA slopes show a range of behaviors.  FIR6B, MMS7, MMS9, and NW167 all show SEDs where the ALMA and NOEMA slopes are in agreement, indicating that the two instruments are tracing the same continuum emission mechanism. By contrast, FIR2 and MMS6 have discrepant ($>2 \sigma$) slopes between the ALMA and NOEMA data, suggesting that they are tracing different types of emission. 

    \item Free-free emission from bremsstrahlung and anomalous microwave emission (AME) are both likely insufficient to explain the change in slope. 

    \item 
    Contamination from disk emission could be up to 70\% in FIR6B and MMS7, but the disk contributions were $\lesssim$ 40\% in FIR2, MMS6, MMS9, and NW167.  Only for FIR2 could we rule out disk contamination.  Therefore, the flattened slopes could be explained by contributions from disk emission for some protostars but not for all of them. 

    \item  Our analysis suggests that disk contamination could be a non-negligible factor on envelope and core scales, including high resolution single-dish telescopes at low frequencies (e.g., GBT).  Multi-wavelength studies spanning a range of spatial scales and frequencies may need to take into account multiple dust components to accurately infer the SED shape and measure dust properties.
    
\end{enumerate}

Future work will be necessary to understand better the SEDs of these sources and to model their grain properties.  While this study focuses specifically on Orion cores and examining individual protostars, \citet{Lowe2022} shows that elevated fluxes at $< 100$ GHz frequencies are seen throughout Orion A, Orion B, and Serpens on cloud (120\arcsec) resolutions.  Therefore, this change in SED slope may arise from emission origins that are on larger scales than disks or protostellar jets.  %it represents an important first step in elucidating the behaviour of dust grains in dense regions of the ISM. The results
Understanding the true shape of the SED over different scales and the origins of any changes in slope are necessary to elucidate the behaviour of dust in the ISM.  Moreover, such models will have broader implications for interpreting dust emission and accurately measuring masses in molecular clouds and cores.

\begin{acknowledgements}
PN and SIS thank the anonymous referee for their comments, which greatly improved the paper, and acknowledge the support provided by the Natural Sciences and Engineering Research Council of Canada (NSERC) through grants RGPIN-2020-03981, RGPIN-2020-03982, and RGPIN-2020-03983. The authors thank the NAASC for support with the ALMA observations and data processing. 
The authors extend their gratitude to Sara Stanchfield, Tony Mroczkowski, Di Li, Zhiyuan Ren, and Lei Zhu for their support and contributions to this work. This paper makes use of the following ALMA data: ADS/JAO.ALMA\#2017.1.00357.S. ALMA is a partnership of ESO (representing its member states), NSF (USA) and NINS (Japan), together with NRC (Canada), MOST and ASIAA (Taiwan), and KASI (Republic of Korea), in cooperation with the Republic of Chile. The Joint ALMA Observatory is operated by ESO, AUI/NRAO and NAOJ. The National Radio Astronomy Observatory is a facility of the National Science Foundation operated under cooperative agreement by Associated Universities, Inc. This work is based on observations carried out under project number W18BG with the IRAM NOEMA Interferometer. IRAM is supported by INSU/CNRS (France), MPG (Germany) and IGN (Spain). 
This research has made use of the SIMBAD database, operated at CDS, Strasbourg, France.  
\end{acknowledgements}

%_______________________________________________________
\bibliography{ref}{}
\bibliographystyle{aasjournal}

\end{document}